\documentclass[
    ,final            
  ]
  {aipproc}

\layoutstyle{8x11double}

\begin{document}

\title{Miras around the Galactic Center}

\classification{<Replace this text with PACS numbers; choose from this list:
                \texttt{http://www.aip..org/pacs/index.html}>}

\author{Noriyuki Matsunaga}{
  address={Institute of Astronomy, University of Tokyo}
}

\author{IRSF/SIRIUS team}{
  address={including Kyoto University, Nagoya University, National Astronomical Observatory of Japan,\\and South African Astronomical Observatory}
}

\begin{abstract}
We report results of our near-IR survey for variables in a field
of view of 20$^\prime$ by 30$^\prime$ towards the Galactic Center (GC),
where we detected 1364 long-period variables. We have established a method
for the simultaneous estimation of distances and extinctions using
the period-luminosity relations for the $JHK_{\mathrm{s}}$ bands.
Our method is applicable to Miras with periods in the range 100--350 days
and mean magnitudes available in two or more filters. Here we discuss
143 Miras whose distances and extinctions were obtained based on their
periods and $H$- and $K_{\mathrm{s}}$-band magnitudes. We find that
almost all of them are located at the same distance to within our accuracy,
and the distance modulus of the GC is estimated to be
$14.58\pm 0.02 \pm0.11$~mag.
The former error corresponds to the statistical error and the latter to
the systematic one which includes the uncertainty of our assumed distance
modulus of the LMC ($18.45\pm 0.05$~mag).
We also discuss the large and highly variable extinction towards the GC.
\end{abstract}

\maketitle


\section{Introduction}

The period-luminosity relation (PLR) of Miras has been widely used
as a distance indicator, such as measuring distances to
nearby galaxies (e.g. \cite{Whitelock-2009}).
We can also investigate the distribution of Miras in the Galaxy using the PLR.
Their important advantage as tracers is that we can obtain a position
of each Mira \cite{Matsunaga-2005}.
In this contribution we present results of our near-IR survey of Miras
towards the GC. Glass and collaborators \cite{Glass-2001} conducted
a K-band (2.2~$\mu$m) survey for almost the same region as ours and
detected 409 long-period variables. Our survey is deeper by more than 
one magnitude and, more importantly, includes the simultaneous monitoring
in $J$ (1.25~$\mu$m) and $H$ (1.63~$\mu$m) as well as $K_{\mathrm{s}}$
(2.14~$\mu$m).

\section{Observation and analysis}

We carried out a variability survey for a field of view of 20$^\prime$
by 30$^\prime$ towards the GC using the IRSF 1.4-m telescope and
the SIRIUS near-IR camera sited at SAAO, Sutherland, South Africa.
Roughly 90 monitoring data were collected between 2001 and 2008. 
The detection limit is approximately 16.4, 14.5 and 13.1~mag
in $J$, $H$, and $K_{\mathrm{s}}$, respectively.
We detected more than 80,000 sources in our field of view,
and we found variations of 1,364 stars. We identified the counterparts
of 347 variables in the catalogue of Glass et al. \cite{Glass-2001}.
Then we found periods for more than 500 variables, a majority of which
are considered to be Miras. The catalogue of variable stars and
their time-series magnitudes will be released in our paper
which has been accepted for MNRAS \cite{Matsunaga-2009}.

\section{Estimation of distance and extinction}

\begin{figure}
  \includegraphics[width=.45\textwidth]{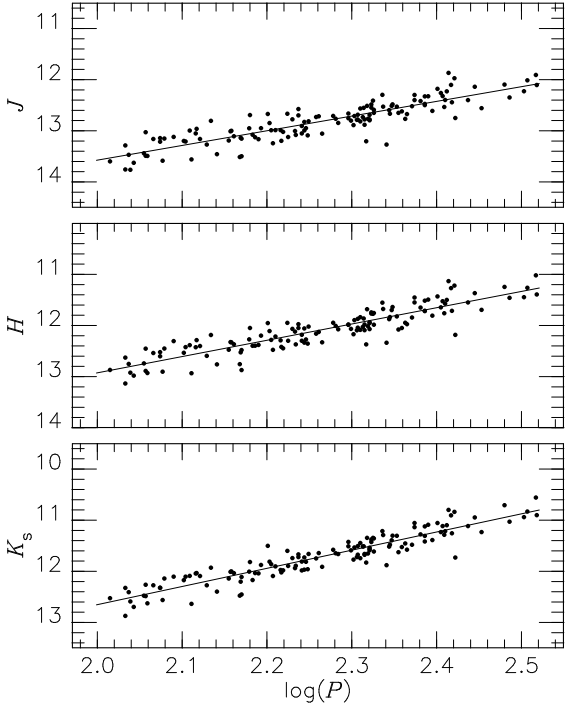}
  \caption{Period-luminosity relation for Miras in the LMC,
plotted for 134 O-rich Miras selected from \cite{Ita-2004}.
\label{fig:LMCPLR}}
\end{figure}

We make use of Miras in the LMC to calibrate the PLR.
134 Miras in the LMC were selected based on their amplitudes,
periods, and colors so as to include oxygen-rich Miras in the period range
of 100--350~days (Fig.~\ref{fig:LMCPLR}).
The period limit is used to exclude objects with
thick circumstellar dust and hot-bottom burning stars.
The data were taken from the catalogue in \cite{Ita-2004}.
The residual scatter around the obtained PLR, the line
in Fig.~\ref{fig:LMCPLR}, is less in $K_{\mathrm{s}}$
(0.17~mag) than in $J$ and $H$ (0.19~mag both), but the differences are small.
This indicates that the PLR in $J$ and $H$ can also be useful for
distance estimation. More importantly, one can also estimate the amount of
interstellar extinction by using the PLR in two or more bands.
We assume that the distance modulus of the LMC is 18.45~mag.
The random error in estimating distances of individual Miras is
around 0.2~mag and the systematic error is around 0.11~mag.
These uncertainties will be discussed in the forthcoming paper
\cite{Matsunaga-2009}.

\section{The distance to the GC}

\begin{figure}
  \includegraphics[width=.45\textwidth]{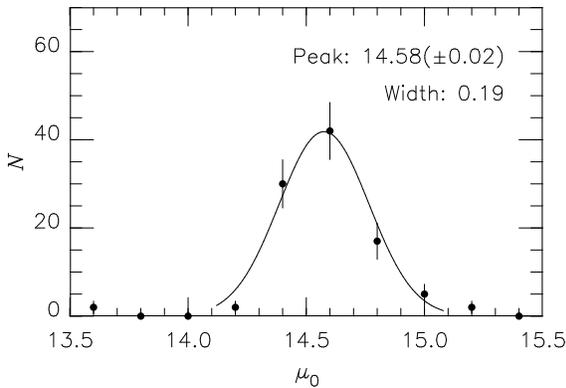}
  \caption{
Distribution of the $\mu_0^{~HK}$ values obtained.
The error bar indicates the size of the Poisson noise ($\sqrt{N}$)
for each bin.
The mean value and width of the fitted Gaussian
are indicated in the panel.
\label{fig:fitDM}}
\end{figure}

While $J$-band magnitudes are often unavailable for our objects
because of the large extinction,
$H$- and $K_{\mathrm{s}}$-band magnitudes are obtained for 143 Miras
and we obtained their distance moduli $\mu_0$
and extinctions $A_{K_{\mathrm{s}}}$.
We estimate the distance to the GC based on 100 Miras which have
values of $14<\mu_0< 15$ and $1<A_{K_{\mathrm{s}}} <3$ to exclude
the selection bias. As shown in Fig.~\ref{fig:fitDM},
the distribution is well fitted with a Gaussian function.
Thus $\mu_0(\mathrm{GC})$ is estimated at $14.58\pm 0.02\pm 0.11$~mag including
the systematic error. The statistical error is small while
the systematic one is significant. This result agrees
with recent results (e.g. \cite{Gillessen-2009}) within the uncertainty.

\section{Interstellar extinction}

Because we can estimate distances and
extinctions for individual Miras, they can be used as tracers to
reveal three-dimensional structure. It is possible to study how the
extinction changes with distance along various lines of sight, although the
low space density of Miras may prevent us from investigating small
structures.

\begin{figure*}
  \includegraphics[width=.85\textwidth]{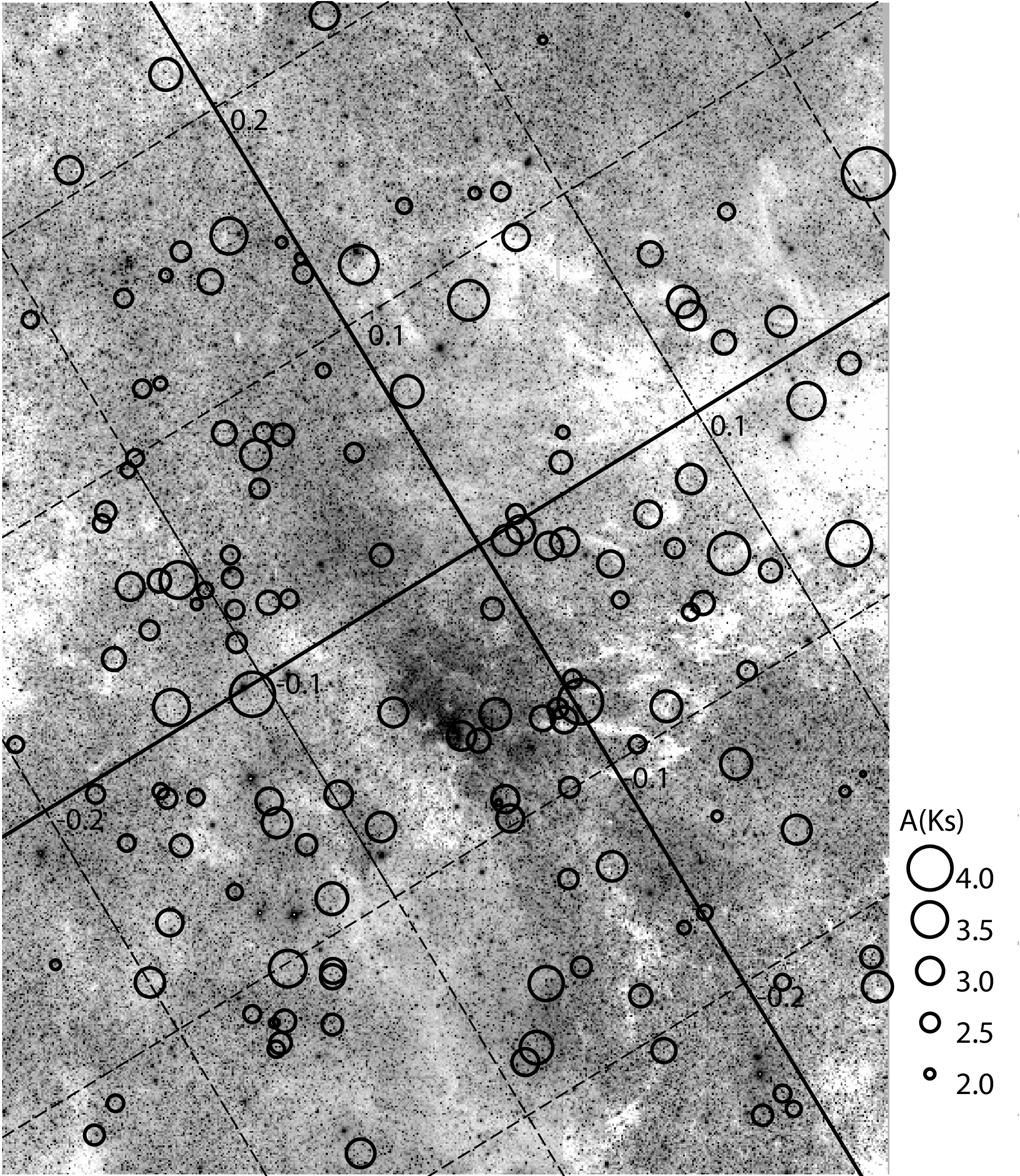}
  \caption{
Locations of 143 Miras with $A_{K_{\mathrm{s}}}$ obtained,
an $H$-band image of the observed field.
North in the equatorial system is up, and east is to the left.
The obtained extinction for each object is indicated by the size of a circle
as shown besides the panel.
\label{fig:AKchart}}
\end{figure*}

In spite that our objects are located at the same distance,
the obtained extinctions cover a large range;
from 1.5~mag to larger than 4~mag in
$A_{K_{\mathrm{s}}}$ except towards the thicker dark nebulae.
It also varies in a complicated way with the line of sight. 
In Fig.~\ref{fig:AKchart}, we plot the locations of the Miras
with extinction values in the galactic coordinate.
The sizes of the symbols vary according to $A_{K_{\mathrm{s}}}$ as indicated.

\section{Concluding remarks}

We have confirmed a method for obtaining distances and extinctions
for Miras with periods between 100 and 350~d by making use of their PLR.
The technique depends on knowing mean magnitudes in two or more filters
in the near infrared. We estimated the distance of the GC to be 
$8.24\pm 0.08~({\it stat.}) \pm 0.42~({\it syst.})$~kpc.
The PLR of Miras is
shown to be a powerful tool for studying Galactic structure, although there
remains a significant systematic error. It is expected that better
calibrations of the PLR will improve the accuracy of the method; this may be
achieved by parallax projects such as VERA and GAIA.

\begin{theacknowledgments}
NM acknowledges that a part of the travel expense to Santa Fe
was supported by Hayakawa Yukio Fund operated by
the Astronomical Society of Japan.
\end{theacknowledgments}

\bibliographystyle{aipprocl} 

\end{document}